\documentclass{article}

\usepackage{booktabs}
\usepackage[utf8]{inputenc}
\usepackage{biblatex}
\usepackage[T1]{fontenc} % T2A for cyrillics
\usepackage[fleqn]{amsmath}
\usepackage{graphicx}
\usepackage{siunitx}
\usepackage{enumitem}
\usepackage{csquotes}
\usepackage{xurl}
\usepackage{lipsum}
\usepackage[most]{tcolorbox}
\usepackage{tabularx}
\usepackage{multirow}
\usepackage{subcaption}
\usepackage{rotating}
\usepackage{amsthm}
\usepackage{sansmath}
\usepackage[table]{xcolor}
\usepackage{xspace}
\usepackage[acronym]{glossaries}
\usepackage{cleveref}

\newacronym[description={Fourth Generation of Cellular Network Technology}]{4G}{4G}{fourth generation of cellular network technology}
\newacronym[description={Fifth-Generation Technology Standard for Cellular Networks}]{5G}{5G}{fifth-generation technology standard for cellular networks}
\newacronym[description={Sixth-Generation Technology Standard for Cellular Networks}]{6G}{6G}{sixth-generation technology standard for cellular networks}
\newacronym[description={Access and Mobility Management Function}]{AMF}{AMF}{access and mobility management function}
\newacronym[description={Application Programming Interface}]{API}{API}{application programming interface}
\newacronym[description={Autonomous System}]{AS}{AS}{autonomous system}
\newacronym[description={Affinity for Technology Interaction}]{ATI}{ATI}{affinity for technology interaction}
\newacronym[description={Authentication Server Function}]{AUSF}{AUSF}{authentication server function}
\newacronym[description={Content Delivery Network}]{CDN}{CDN}{content delivery network}
\newacronym[description={Commercial Internet Exchange}]{CIX}{CIX}{commercial Internet exchange}
\newacronym[description={Domain Name System}]{DNS}{DNS}{domain name system}
\newacronym[description={Digital Object Identifier}]{DOI}{DOI}{digital object identifier}
\newacronym[description={Disruption-Tolerant Network}]{DTN}{DTN}{disruption-tolerant network}
\newacronym[description={Edge Application Server Discovery Function}]{EASDF}{EASDF}{edge application server discovery function}
\newacronym[description={Integrated Access and Backhaul}]{IAB}{IAB}{integrated access and backhaul}
\newacronym[description={Connectivity Monitoring Function}]{CMF}{CMF}{connectivity monitoring function}
\newacronym[description={Journal Impact Factor}]{JIF}{JIF}{journal impact factor}
\newacronym[description={Information and Communication Technology},longplural={information and communication technologies}]{ICT}{ICT}{information and communication technology}
\newacronym[description={IP Multimedia Subsystem}]{IMS}{IMS}{IP multimedia subsystem}
\newacronym[description={Isolated Operation for Public Safety}]{IOPS}{IOPS}{isolated operation for public safety}
\newacronym[description={Inter-Quartile Range}]{IQR}{IQR}{inter-quartile range}
\newacronym[description={Instructed Response Item}]{IRI}{IRI}{instructed response item}
\newacronym[description={Internet Service Provider}]{ISP}{ISP}{Internet service provider}
\newacronym[description={Internet Exchange Point}]{IXP}{IXP}{Internet exchange point}
\newacronym[description={Kernel Density Estimation}]{KDE}{KDE}{kernel density estimation}
\newacronym[description={Multi-Access Edge Computing}]{MEC}{MEC}{multi-access edge computing}
\newacronym[description={Mobile Network Operator}]{MNO}{MNO}{mobile network operator}
\newacronym[description={Network Exposure Function}]{NEF}{NEF}{network exposure function}
\newacronym[description={Network Repository Function}]{NRF}{NRF}{network repository function}
\newacronym[description={Non-Terrestrial Network}]{NTN}{NTN}{non-terrestrial network}
\newacronym[description={Policy Control Function}]{PCF}{PCF}{policy control function}
\newacronym[description={Point of Presence},longplural={points of presence}]{PoP}{PoP}{point of presence}
\newacronym[description={Radio Access Network}]{RAN}{RAN}{radio access network}
\newacronym[description={Session Management Function}]{SMF}{SMF}{session management function}
\newacronym[description={Subscriber Identity Module}]{SIM}{SIM}{subscriber identity module}
\newacronym[description={Self-Reported Single Item}]{SRSI}{SRSI}{self-reported single item}
\newacronym[description={State Synchronization Function}]{SSF}{SSF}{state synchronization function}
\newacronym[description={Uncrewed Aerial Vehicle}]{UAV}{UAV}{uncrewed aerial vehicle}
\newacronym[description={Unified Data Repository}]{UDR}{UDR}{unified data repository}
\newacronym[description={Unified Data Management}]{UDM}{UDM}{unified data management}
\newacronym[description={User Plane Function}]{UPF}{UPF}{user plane function}

\newacronym{3GPP}{3GPP}{3rd Generation Partnership Project}
\newacronym{5G-A}{5G-A}{5G-Advanced}
\newacronym{5G-NR}{5G-NR}{5G New Radio}
\newacronym{ACM}{ACM}{Association for Computing Machinery}
\newacronym{BGP}{BGP}{Border Gateway Protocol}
\newacronym{CSCW}{CSCW}{Computer-Supported Collaborative Work}
\newacronym{DoS}{DoS}{Denial-of-Service}
\newacronym{GPS}{GPS}{Global Positioning System}
\newacronym{HCI}{HCI}{Human-Computer Interaction}
\newacronym{ICMP}{ICMP}{Internet Control Message Protocol}
\newacronym{ICORE}{ICORE}{International CORE Conference Rankings}
\newacronym{iGDB}{iGDB}{Internet Geographic Database}
\newacronym{IoT}{IoT}{Internet of Things}
\newacronym{IP}{IP}{Internet Protocol}
\newacronym{ISO}{ISO}{International Organization for Standardization}
\newacronym{ITU-T}{ITU-T}{International Telecommunication Union Telecommunication Standardization Sector}
\newacronym{NIST}{NIST}{National Institute of Standards and Technology}
\newacronym{O-RAN}{O-RAN}{Open Radio Access Network}
\newacronym{RRC}{RRC}{Radio Resource Control}
\newacronym{VoIP}{VoIP}{Voice over Internet Protocol}
\newacronym{WWW}{WWW}{World Wide Web}

\usepackage{framed} % for shaded env's
\usepackage{stix} % for the black hexagon symbol

\definecolor{mybeige}{HTML}{DDC1B1}
\definecolor{myblue}{HTML}{5E62B8}
\definecolor{mypurple}{HTML}{C5C6E7}
\definecolor{mygray}{HTML}{878697}
\definecolor{mylightbeige}{HTML}{F5DCC9}
\definecolor{mylightpurple}{HTML}{E8E6FE}
\colorlet{shadecolor}{mybeige!30}
\colorlet{altrowcolor}{myblue!30}

\newcolumntype{R}{>{\raggedleft\arraybackslash}X}
\newcolumntype{L}[1]{>{\raggedright\arraybackslash}p{#1}}
\newcolumntype{C}[1]{>{\centering\arraybackslash}p{#1}}

\newcommand{\tfmark}[1]{\textsuperscript{#1}}
\newcommand{\tfnote}[2]{\tfmark{#1} #2}
\newcommand{\tfnotes}[3]{\multicolumn{#1}{p{#2}}{\tiny{#3}} \\ \tiny{} \\}

\DeclareSIUnit{\year}{years}

\newcommand{\full}{\emx{\blacksquare}}
\newcommand{\half}{\emx{\squareulblack}}
\newcommand{\emty}{\emx{\square}}

\newcommand{\ctxtriangle}{\emx{\blacktriangle}}
\newcommand{\ctxcircle}{\emx{\mdsmblkcircle}}
\newcommand{\ctxsquare}{\emx{\mdsmblksquare}}
\newcommand{\ctxhexagon}{\emx{\hexagonblack}}

\newcommand{\strongemph}[1]{{\textbf{#1}}}

\newcommand{\hlblue}[1]{\setlength{\fboxsep}{1pt}\colorbox{myblue!30}{#1}}
\newcommand{\hlgray}[1]{\setlength{\fboxsep}{1pt}\colorbox{mygray!30}{#1}}

\newcommand{\captionstrongemph}[1]{{\textbf{#1}}}
\newcommand{\captionheadline}[1]{\captionstrongemph{#1.}\enspace}

\newcommand{\numgnbs}{310\xspace} % 176 in Paris and 134 in Lyon
\newcommand{\numtiles}{\num{1871}\xspace} % 934 in Paris and 937 in Lyon

\newcommand{\eg}{e.\,g.\xspace}
\newcommand{\ie}{i.\,e.\xspace}

\newcommand{\emx}[1]{\ensuremath{#1}\xspace}

\newtheoremstyle{researchquestionstyle}
  {0pt}                                      % Space above
  {0pt}                                      % Space below
  {\normalfont}                              % Body font
  {0pt}                                      % Indent
  {\normalfont\normalsize\textbf}   % Heading font
  {.}                                        % Punctuation
  {0.5em}                                    % Space after heading
  {}                                         % Default heading format
\theoremstyle{researchquestionstyle}
\newtheorem{rqtheorem}{Research Question}
\crefname{rqtheorem}{RQ}{RQs}
\Crefname{rqtheorem}{RQ}{RQs}

\newlist{outlineitems}{itemize}{1}
\setlist[outlineitems]{
  label=\textcolor{red}{$\bullet$},
  leftmargin=*,
  before=\color{red},
}

\newenvironment{emphbox}
{%
  \begin{shaded}
  \noindent
}
{%
  \end{shaded}
}

\begin{document}

\title{Island-Ready Communication Networks: Connectivity Monitoring and State Synchronization for 5G Crisis Connectivity at the Edge}

\author{%
    Leon Janzen\textsuperscript{*}, Maria Koch\textsuperscript{*}, Matthias Hollick\textsuperscript{*,$\dagger$} \\
    \textit{\textsuperscript{*}Secure Mobile Networking Lab} \\
    Technical University of Darmstadt \\
    Darmstadt, Germany \\
    \{ljanzen,makoch,mhollick\}@seemoo.tu-darmstadt.de \\
    \textit{\textsuperscript{$\dagger$}Secure Mobile Systems Group} \\
    IMDEA Networks Institute \\
    Madrid, Spain \\
    matthias.hollick@networks.imdea.org 
}

\maketitle
% \setlength{\fboxrule}{2pt}
% \begin{tikzpicture}[remember picture,overlay]
% \node[anchor=south,yshift=10pt] at (current page.south) {\fcolorbox{red}{white}{\parbox{\dimexpr0.65\textwidth-\fboxsep-\fboxrule\relax}{\footnotesize \textbf{Draft.} This work will be submitted to the IEEE for possible publication. Copyright may be transferred without notice, after which this version may no longer be accessible.}}};
% \end{tikzpicture}

% ################################################################################
% ################################################################################
% ################################################################################
% ################################################################################
% ################################################################################

\glsunset{5G}

\begin{abstract}
Island-ready \gls{5G} deployments provide edge connectivity for isolated regions in crisis scenarios by switching to island operation when they are cut off from the global Internet.
This work addresses open research challenges that hinder the practical realization of island readiness, with a focus on the required state synchronization and connectivity monitoring functionality.
As a proof of concept and for performance evaluation, we integrate our solutions in open-source projects and emulate a country-wide \gls{5G} communication network with real-world cellular traffic.
Our evaluation studies the switching latency of transitions to island operation, showing that most of the transition delay stems from the access network reconfiguration.
We study the synchronization overhead of a highly distributed core network, finding that the synchronization traffic grows faster than linearly with the number of \gls{5G} cores, reaching peak synchronization data rates of \SI{9.36}{\mega\byte\per\second} for a deployment of 50 cores.
\end{abstract}

% \begin{keywords}
% \input{keywords}
% \end{keywords}

% ################################################################################
% ################################################################################
% ################################################################################
% ################################################################################
% ################################################################################

\section{Introduction}

\gls{5G} mobile networks are a complex system of systems, whose availability is taken for granted by the general public.
However, the architecture of traditional \gls{5G} deployments, with only a few control-plane sites per country, results in long backhaul distances that are susceptible to failure.
On multiple occasions during recent crises, regions were cut off from the \gls{5G} network as a result of natural disasters or armed conflicts~\cite{fekete_2021_here,jain_2022_ukrainian,guilfoyle_2022_final,internetsociety_2024_west,gonzales_2025_social}.
In the aftermath, survivors could not rely on \gls{5G} connectivity during crisis response.

The concept of island readiness~\cite{a-visionpaper} aims to build resilience against such isolation incidents by distributing the \gls{5G} control plane across a high number of regions.
While prior work demonstrates a minimal connectivity and synchronization proof of concept for \gls{5G} cores with focus on end-to-end connectivity and real smartphones~\cite{c-decentralcore}, no large-scale evaluation exists.
In addition, open challenges remain around transitions between normal and island operation and the systematic integration of synchronization and connectivity functionality in the \gls{5G} control plane.
Our first research question addresses the state synchronization and connectivity monitoring:

\begin{emphbox}
\textbf{RQ1.}
How can the state synchronization and connectivity monitoring of island-ready \gls{5G} cores benefit from failure detection and consensus protocols?
\end{emphbox}

\emph{Connectivity monitoring} is necessary to detect isolation from global connectivity, triggering the activation of regional failover cores into island operation.
In island operation, \gls{5G} connectivity is provided in a regional scope.
This enables users to reach services hosted at the edge through the \gls{5G} access network, \eg, applications relevant for crisis response.
Continuous \emph{state synchronization} is necessary for two reasons:
It enables distributed \gls{5G} cores to maintain a common global system view, and it enables failover cores to take over based on the most recent state.
There exist numerous related works in the realms of failure detection and consensus protocols, which we review in \Cref{sec: pub-d related work} regarding their applicability to connectivity monitoring and state synchronization.
Based on these, we introduce two novel \gls{5G} network functions in \Cref{sec: pub-d design}, the \gls{SSF} and the \gls{CMF}, that make state-of-the-art \gls{5G} deployments island-ready.
We design and implement the \gls{SSF} and \gls{CMF} as custom network functions in the control plane of the popular open-source \gls{5G} core implementation \emph{Open5GS}~\cite{open5gs}.

The concept of island readiness promises improved \gls{5G} connectivity, as it enables edge connectivity in isolated regions that would otherwise not be supported.
However, it introduces overhead to the \gls{5G} system, as state synchronization and connectivity monitoring require additional resources.
Our second research question is therefore concerned with the resource consumption of the novel \gls{SSF} and \gls{CMF}:

\begin{emphbox}
\textbf{RQ2.}
Which overhead do state synchronization and connectivity monitoring add to country-wide island-ready \gls{5G} deployments?
\end{emphbox}

\Glspl{MNO} typically deploy country-wide \gls{5G} systems, which invites evaluation of island readiness in that setting.
To this end, we derive a system model in \Cref{sec: pub-d system model} by generating topologies for country-wide \gls{5G} access and core networks with \emph{POBTOG}~\cite{poignee_2025_pobtog}.
We fine-tune parameters to align our system model with the real world and integrate the \emph{NetMob} dataset~\cite{martinez_2023_netmob} to simulate user uplink and downlink traffic to application servers via the \gls{5G} user plane.
\Cref{sec: pub-d setup} describes our 14-node Kubernetes cluster, on which we orchestrate an island-ready \gls{5G} control plane for France and simulate user plane connections of users in Paris and Lyon.
Our large-scale simulation study allows evaluations of island readiness in a realistic setting.
\Cref{sec: pub-d evaluation} covers our evaluation of the switching latency and synchronization overhead of the novel \gls{SSF} and \gls{CMF}.
Our findings indicate that the synchronization overhead of island-ready communication networks grows faster than linearly with the number of \gls{5G} cores.

In summary, we solve open challenges of island readiness by integrating \gls{SSF} and \gls{CMF} network functions in the \gls{5G} control plane (\Cref{sec: pub-d related work,sec: pub-d design}).
We conduct a large-scale simulation study of country-wide communication networks (\Cref{sec: pub-d system model,sec: pub-d setup}) and evaluate the performance and resource overhead of island readiness (\Cref{sec: pub-d evaluation}).
With these contributions, we aim to accelerate the realization of island-ready \gls{5G} and beyond mobile networks.
\emph{We publish the Ansible playbooks, Helm charts, Kubernetes manifest files, Docker images, and Go scripts to allow reproducing and extending our findings.}

% ################################################################################
% ################################################################################
% ################################################################################
% ################################################################################
% ################################################################################

\section{Background and Related Work}
\label{sec: pub-d related work}

The centerpiece of \gls{5G} communication networks is the \gls{5G} core, which is separated into a control and a user plane~\cite{3gpp-23-501}.
With \gls{5G-A}, the \gls{UPF} can be deployed in a distributed manner to guide user traffic to the data network more directly.
Still, the initial connection establishment has to be routed through the remote control plane, \eg, to the \gls{AMF}.
Today's typical \gls{5G} deployments are assumed to feature a handful of \gls{5G} control-plane locations per country that provide a high-availability backend, minimizing the probability that all \gls{5G} cores are unavailable at the same time~\cite{rula_2017_adopting}.
However, the long backhaul distances between users and the closest \gls{5G} core remain a major weakness in \gls{5G} deployments.
Recent events have shown that natural disasters and armed conflicts likely cause Internet outages as a side effect.
As a result, crisis response is often hindered because survivors cannot use \gls{5G} connectivity to coordinate efforts.

\subsection{Island Readiness}

The concept of island connectivity~\cite{b-userperspective} describes how users can benefit from regional connectivity when global connectivity breaks, for example, to maintain communication, navigation, and information services for survivors during crisis response.
State-of-the-art connectivity solutions for isolation scenarios include mobile \gls{5G} cores with integrated base stations that can be brought to affected regions~\cite{lukau_2024_public}.
However, setting up these solutions takes time, which can be crucial during crisis response, while island connectivity seamlessly maintains regional connectivity.
The vision of island readiness~\cite{a-visionpaper} analyzes which actors play a role in making \gls{5G} networks island-ready, \eg, \glspl{MNO} distributing the \gls{5G} control plane, application developers adapting regional-first server architectures, and cloud providers creating additional edge capacity to keep up with changing demands.
Today's \gls{5G} deployments can be made island-ready by adding connectivity and synchronization functionality to state-of-the-art \gls{5G} control planes, and related work demonstrates the basic functionality of these components in a minimal lab setup~\cite{c-decentralcore}.
This work aims to support the ongoing quest for island readiness by filling the following research gaps:

\begin{emphbox}
    \textbf{Research Gap.}
    First, there has been no extensive survey of failure detection and consensus protocols for the connectivity and synchronization functionality of island-ready \gls{5G} cores (RQ1).
    Second, there has been no extensive evaluation of the overhead introduced by technologies enabling island readiness (RQ2).
\end{emphbox}

\subsection{Related Work}

\paragraph{Consensus Protocols}

Over the past few decades, a broad spectrum of distributed consensus algorithms has emerged under different failure assumptions.
There are crash fault tolerance protocols, such as \emph{Paxos}~\cite{lamport_1998_paxos} and \emph{Raft}~\cite{ongaro_2014_raft}, which evolved into \emph{Pirogue}~\cite{paris_2015_pirogue} and \emph{RaftOptima}~\cite{kondru_2024_raftoptima}.
In addition, there are Byzantine fault tolerance protocols, such as \emph{PBFT}~\cite{castro_2002_pbft}, \emph{Tendermint}~\cite{kwon_2014_tendermint}, and \emph{HotStuff}~\cite{yin_2019_hotstuff}, which evolved into \emph{Bullshark}~\cite{spiegelman_2022_bullshark} as well as \emph{Narwhal and Tusk}~\cite{danezis_2022_narwhaltusk}.

\paragraph{Failure Detection}

Basic failure detection is often implemented using health-check protocols, where components regularly exchange short heartbeat-like messages~\cite{kawazoe_1997_heartbeat}.
Beyond that, Chandra et al. laid the basis for more advanced failure detectors~\cite{chandra_1996_unreliable}, proving it impossible to design a failure detector that holds strong completeness and perfect accuracy in asynchronous environments.
The $\phi$-accrual failure detector~\cite{hayashibara_2004_phi} offers high scalability through statistical analysis of individual links.
It was extended into a protocol for large-scale distributed systems~\cite{lavinia_2010_failure}.
\emph{Gossip}~\cite{renesse_2009_gossip} introduced epidemic-style heartbeat checks with random subsets of neighbors, which influenced \emph{SWIM}~\cite{das_2002_swim}, \emph{Lifeguard}~\cite{dadgar_2018_lifeguard}, and \emph{memberlist}~\cite{hashicorp_2025_memberlist} that is used in \emph{Serf}~\cite{phillips_2025_serf} and \emph{Consul}~\cite{hashicorp_2025_consul}.
\emph{Falcon}~\cite{leners_2011_falcon} aims to design a reliable failure detector by coordinating a network of spies monitoring different layers.

\paragraph{Related 5G Simulation Studies}
Numerous papers describe experiments with cloud-native \gls{5G} deployments in Kubernetes~\cite{barrachina_2022_cloud,wijesekera_2025_kube5gc}, some of which consider the regional aspects of \gls{MEC}~\cite{ungureanu_2021_collaborative}.
Others discuss the availability and resilience of \gls{5G} deployments~\cite{simone_2024_performance,paul_2025_scalable}.
There are performance studies with large-scale \gls{5G} simulations~\cite{subramanian_2025_performance}.
\emph{CoreKube}~\cite{larrea_2023_corekube} presents a cloud-native resilient \gls{5G} core design that focuses on scalability instead of transitions to regional connectivity.
Island-ready \gls{5G} deployments have only been simulated in minimal form, and 
\strongemph{this work presents the first large-scale simulation study of island-ready communication networks}.

% ################################################################################
% ################################################################################
% ################################################################################
% ################################################################################
% ################################################################################

\section{Design}
\label{sec: pub-d design}

\begin{figure}
    \includegraphics[width=\linewidth]{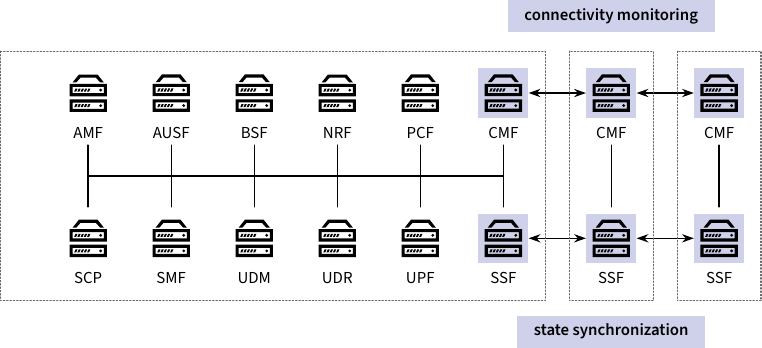}
    \caption[Island-Ready 5G Control Plane]{%
        \captionheadline{Island-Ready \Gls{5G} Control Plane}
        The boxes mark three \gls{5G} cores from different regions.
        We add \hlblue{novel \gls{CMF} and \gls{SSF} network functions} to make \gls{5G} cores island-ready.
        The remaining network functions are explained in TS 23.501~\cite{3gpp-23-501}.
    }
    \label{fig: pub-d control plane}
\end{figure}

We integrate state synchronization and connectivity monitoring into island-ready \gls{5G} cores by adding novel \gls{CMF} and \gls{SSF} network functions to the \gls{5G} control plane.
\Cref{fig: pub-d control plane} visualizes the interplay of these network functions across multiple \gls{5G} cores.

\subsection{Connectivity Monitoring Function (CMF)}
The \gls{CMF} monitors connectivity between \gls{5G} cores.
This connection is typically preconfigured, \eg, for many regional failover cores monitoring the same active core, which drastically simplifies the task description.
As a result, the benefits of advanced failure detection protocols do not apply, and the additional overhead they incur makes them undesirable to use.
Instead, we implement the \gls{CMF} with a heartbeat-based protocol that checks the liveness of other \glspl{CMF}, triggering transitions between normal and island operation based on failure and recovery thresholds.

\subsection{State Synchronization Function (SSF)}

\begin{table}
  \setlength{\tabcolsep}{4pt}
  \caption[Consensus Protocol Survey]{
    \captionheadline{Consensus Protocol Survey}
    Protocols fulfill criteria fully (\full), partially (\half), or not at all (\emty).
  }
  \label{tab: pub-d consensus}
  \begin{tabularx}{\linewidth}{Xccccc}
    \toprule % ------------------------------------------------------------------
    \textbf{Protocol} &
    \textbf{Applicability} &
    \textbf{Availability} &
    \textbf{Consistency} &
    \textbf{Recovery} \\
    \midrule % ------------------------------------------------------------------
    SWIM~\cite{das_2002_swim}                  & \full & \full & \emty & \emty \\
    $\phi$ Accrual~\cite{hayashibara_2004_phi} & \full & \full & \emty & \emty \\
    \rowcolor{myblue!30}
    Raft~\cite{ongaro_2014_raft}               & \full & \full & \full & \full \\
    Pirogue~\cite{paris_2015_pirogue}          & \half & \emty & \full & \full \\
    RaftOptima~\cite{kondru_2024_raftoptima}   & \half & \emty & \full & \full \\
    Falcon~\cite{leners_2011_falcon}           & \emty & \half & \emty & \emty \\
    Paxos~\cite{lamport_1998_paxos}            & \emty & \emty & \full & \full \\
    PBFT~\cite{castro_2002_pbft}               & \emty & \emty & \full & \full \\
    \bottomrule % ---------------------------------------------------------------
  \end{tabularx}
\end{table}

The \gls{SSF} replicates the state of all \gls{5G} network functions between the cores.
This enables failover cores to activate in island operation and serve users with island connectivity based on the most recent state.
In addition, the \gls{SSF} must handle divergent states, \eg, when failover cores reconnect to the core network after an outage.
For these tasks, suitable consensus protocols need to handle state consistency across the nationwide deployment.
\Cref{tab: pub-d consensus} summarizes our consensus protocol survey regarding these criteria, including the protocol's applicability to island readiness and the availability of its source code.
\emph{Raft}~\cite{ongaro_2014_raft} achieves consensus efficiently with a small subset of voting members that elect a primary for write operations.
% The combination of a consensus protocol and the replication protocol based on a single primary member, suggest \emph{Raft} for the implementation of the \gls{SSF}.

% ################################################################################
% ################################################################################
% ################################################################################
% ################################################################################
% ################################################################################

\section{System Model}
\label{sec: pub-d system model}

Our system model covers end-to-end connections between users and application servers through the \gls{5G} network.

\paragraph{Country-Wide 5G Deployments}
POBTOG~\cite{poignee_2025_pobtog} creates country-wide communication network topologies using population density as a proxy for the number of end users per area.
POBTOG generates hierarchical networks of $n$ layers, requiring $n=4$ layers to model raft cores, active cores, failover cores, and base stations.
The topology and node count per layer are configured through the parameters $T$ and $N$.
The number of nodes of the lowest layer, $N_{n-1}$, is automatically determined through parameter $P$, as POBTOG iteratively divides the configured country into Voronoi cells until each cell holds no more than $P$ inhabitants.

\paragraph{End-to-End Connectivity for Users and Applications}

In our model, applications are hosted on the Internet and accessed through \gls{5G} connectivity.
The user equipment, \ie, the user's smartphone, connects to a nearby base station, adding a star-topology access layer below the POBTOG network topology.
Users send uplink traffic to the application servers, and receive downlink traffic in return.
We model the application server deployment that maps each application to regions with local access after selecting two disjoint subsets of traditionally hosted and island-ready applications.
In this context, a \emph{region} means the area served by an active or failover \gls{5G} core, and island-ready applications feature servers in all regions.

% \paragraph{The Locality of Infrastructure}
% The goal of island connectivity is to maintain regional connectivity when global connectivity fails.
% To determine which systems are available and which are not during isolation challenges, the location of all systems has to be considered.
% For instance, regions could be determined at the hand of a POBTOG layer, such as dividing the country into $N_2$ regions, where each region contains the infrastructure (\ie, of all nodes below one of the $L_2$ failover cores).

\paragraph{Single Snapshots vs. Dynamic Simulation}
The aforementioned notations enable modeling end-to-end connections between users and application servers via \gls{5G} for a single snapshot of the system.
Modeling the system over time requires considering timeslots and the uplink and downlink traffic per timeslot.
% Challenges determine which subsystems are unavailable during which timeslots, allowing us to model regional failures and simulate the transition of the \gls{5G} deployment into and out of island operation.

\begin{table}    
  \caption[System Model]{
    \captionheadline{System Model}
    We model a \Gls{5G}-based communication network (POBTOG) and emulate end-to-end connectivity to messenger applications with real-world cellular traffic (NetMob).
  }
  \label{tab: pub-d system model}
  \begin{tabularx}{\linewidth}{lXllS[table-format=6]}
    \toprule % -------------------------------------------------------------------------------------------------------------
    \textbf{System} & \textbf{Layer}          & \textbf{Source}                    & \textbf{Topology} & \textbf{\#Nodes} \\
    \midrule % -------------------------------------------------------------------------------------------------------------
    Data            & Application deployments & NetMob~\cite{martinez_2023_netmob} & Mesh              & 6                \\
    \midrule % -------------------------------------------------------------------------------------------------------------
                    & Raft cores              &                                    & Mesh              & 5                \\
    \gls{5G} Core   & Active cores            & POBTOG~\cite{poignee_2025_pobtog}  & Star              & 25               \\
                    & Failover cores          &                                    & Star              & 500              \\
    \midrule % -------------------------------------------------------------------------------------------------------------
    \gls{5G} RAN    & Base stations           & POBTOG~\cite{poignee_2025_pobtog}  & Ring              & 64752            \\
    \midrule % -------------------------------------------------------------------------------------------------------------
    Users           & User equipment          & NetMob~\cite{martinez_2023_netmob} & Star              & 876593\tfmark{a} \\
    \bottomrule % ----------------------------------------------------------------------------------------------------------
    \tfnotes{5}{\linewidth}{
      \tfnote{a}{One node corresponds to the users in one $100\times100$ m NetMob tile.}
      }
  \end{tabularx}
\end{table}

\subsection{Practical Considerations}
We configure our system model parameters to simulate a realistic country-wide \gls{5G} deployment, combining available datasets as summarized in \Cref{tab: pub-d system model}.
Where applicable, we highlight the limitations of our design choices and mention alternative approaches.
First of all, we choose to simulate a \gls{5G} deployment in France, which is mainly concentrated around Paris but also includes metropolitan areas, medium-sized cities, and rural areas.

\paragraph{5G Core and Access Network Deployments}
Traditional deployments are expected to feature three to five \gls{5G} cores for countries the size of France.
These cores are typically deployed in a high-availability setup spanning multiple regions.
Since our consensus protocols survey suggests Raft-based synchronization, and Raft is designed for three, five, or seven voting members, we simulate $N_0=5$ raft cores.
Our work is motivated by the goal of increasing the end-to-end resilience of \gls{5G} deployments by adding active and failover cores to achieve higher geo-redundancy and enable more regions to provide island connectivity.
Accordingly, we use Level $L_1$ to model active cores, \ie, \gls{5G} cores that are operating non-stop in addition to the raft cores.
Level $L_2$ models failover cores, \ie, \gls{5G} cores that only activate when the region forms an island.
Note that the total number of active and failover cores $N_1+N_2$ determines the number of regions that can form an island upon isolation from the remaining \gls{5G} network.
Related work suggests that failover cores can increase resource efficiency in regions where the permanent operation of an active core is undesirable.
% Level $L_3$ contains the base stations.
% Identifying a suitable parameter $P$ for partitioning France into cells that each resemble a single base station is a separate quest that we discuss next.

\paragraph{Base Stations}

\begin{figure}
   \begin{subfigure}[t]{\linewidth}
      \includegraphics[width=.9\linewidth]{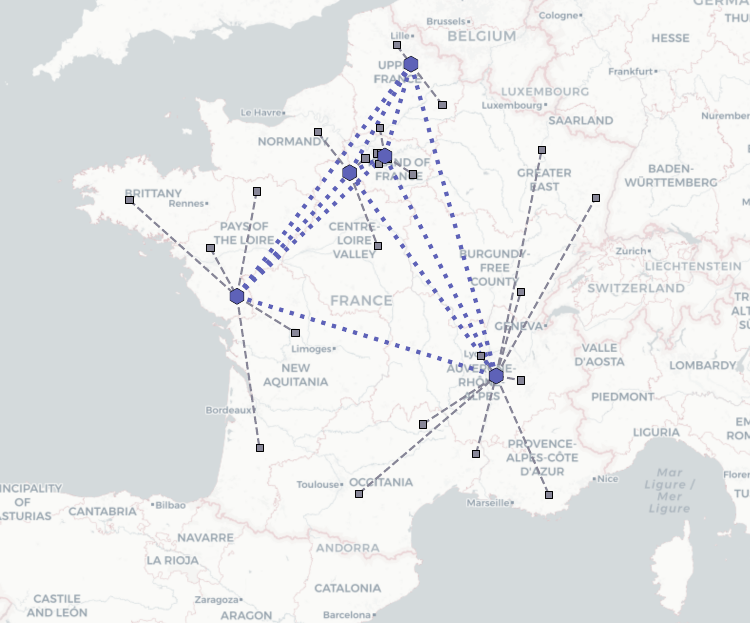}
      \subcaption{Raft and Active Cores}
      \label{fig: pub-d simulation france}
   \end{subfigure}
   \par\vspace{.01\linewidth}
   \begin{subfigure}[t]{.45\linewidth}
      \includegraphics[width=.9\linewidth]{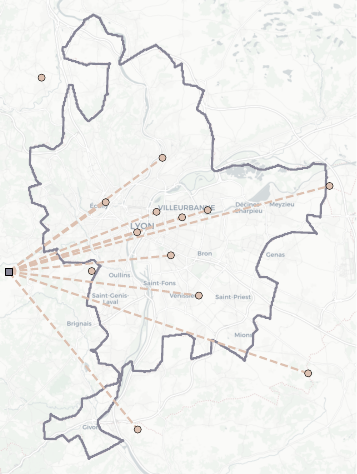}
      \subcaption{Failover Cores}
      \label{fig: pub-d simulation lyon}
   \end{subfigure}
   \hfill
   \begin{subfigure}[t]{.45\linewidth}
      \includegraphics[width=.9\linewidth]{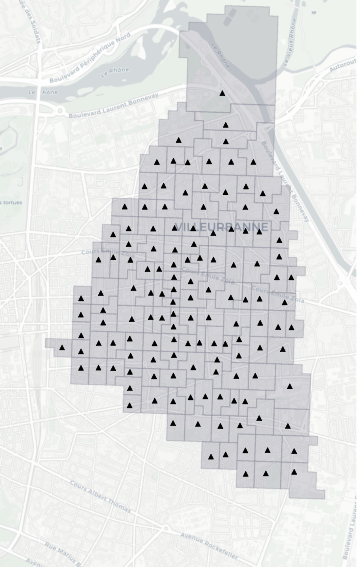}
      \subcaption{Base Stations and NetMob Tiles}
      \label{fig: pub-d simulation region}
   \end{subfigure}
   \caption[Simulation Scope]{%
   \captionheadline{Simulation Scope}
   We simulate five raft cores (\textcolor{myblue}{\ctxhexagon}), one active core (\textcolor{mygray}{\ctxsquare}), two failover cores (\textcolor{mybeige}{\ctxcircle}), \numgnbs base stations (\textcolor{black}{\ctxtriangle}), and the users of \numtiles $100\times100$ m NetMob tiles.
   }
   \label{fig: pub-d france-maps}
\end{figure}

POBTOG has a configuration parameter $P$ that sets the maximum number of users allowed in the cell around a base station.
It iteratively divides the country into Voronoi cells until all cells contain $P$ or fewer users, thereby determining $N_3$.
We test different values of $P$ and compare the resulting number of base stations with the actual number deployed in France.
With $P=1500$, POBTOG generates \SI{64752} base stations, which closely resembles the real number of \SI{63936} base stations reported in~\cite{anfr}.
\Cref{fig: pub-d france-maps} depicts the POBTOG topology with $N=[5,25,500]$, $T=[\text{mesh},\text{star},\text{ring},\text{star}]$, and $P=1500$, showing a France-wide view of the active and raft cores, a close-up of the failover cores in the city of Lyon, and a detailed view of the users of a single region.

\paragraph{Real-World Cellular Traffic}
We aim to incorporate real-world cellular traffic data into our evaluation to achieve results that are grounded in reality.
The NetMob dataset~\cite{martinez_2023_netmob} contains such data for 20 metropolitan areas in France.
It has a spatial resolution of $100\times100$ meter tiles, reporting the combined uplink and downlink traffic for users within that tile to protect the privacy of individual users.
Therefore, we model users within a tile as a single entity and map them to the closest $L_3$ base station.
This mapping of NetMob tiles to POBTOG base stations connects both datasets in the geospatial dimension and extends our system model by a fifth layer.
NetMob reports user traffic for 68 applications, six of which are messengers that a related study found to be a critical application category during island scenarios~\cite{b-userperspective}.
We opt to use messengers as example applications, and simulate a fictitious scenario with two of them:
In our model, WhatsApp is a centrally hosted messenger app, and Telegram is a messenger with an island-ready application backend, causing WhatsApp servers being available in only one region, whereas Telegram servers are deployed across all regions.

\paragraph{Feasible Scope for the Simulation}

The entire network topology includes more than \num{60000} base stations and 530 cores, exceeding our available resources.
To align our simulation scope to our goals, we identified two regions, \texttt{Paris-65080}\footnote{65080 and 64876 are the POBTOG IDs of the \gls{5G} failover cores.} and \texttt{Lyon-64876}, totaling \numtiles NetMob tiles connected to \numgnbs base stations.
We simulate the end-to-end connectivity between these regions and the WhatsApp and Telegram application servers via \Gls{5G}.
% We introduce a third logical region, \texttt{central}, that abstracts any infrastructure deployed outside the Paris and Lyon regions.

% ################################################################################
% ################################################################################
% ################################################################################
% ################################################################################
% ################################################################################

\section{Setup}
\label{sec: pub-d setup}

\subsection{Kubernetes Cluster}

\begin{figure}
    \includegraphics[width=\linewidth]{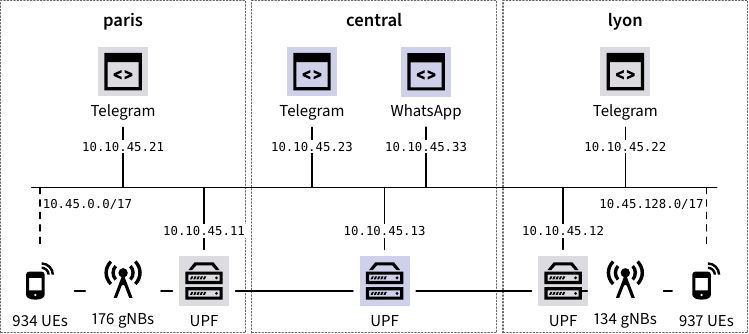}
    \caption[Island-Ready 5G User Plane]{%
    \captionheadline{Island-Ready \gls{5G} User Plane}
    In normal operation, the \hlblue{central \gls{UPF}} connects the \texttt{/16} user equipment network to the global Internet.
    In island operation, the \hlgray{regional \glspl{UPF}} connect their regional \texttt{/17} user equipment network to the regional Internet.
    }
    \label{fig: pub-d user plane}
\end{figure}

We host a Kubernetes cluster on 14 workstations (\emph{nodes}) in a local network, each operating Ubuntu 24.04.
We use \texttt{kubeadm}~\cite{kubeadm} to set up the cluster, \texttt{containerd}~\cite{containerd} as the container runtime, and \emph{Calico}~\cite{calico} as the container networking interface.
As visualized in \Cref{fig: pub-d user plane}, the application servers and \glspl{UPF} share a separate data network deployed with the \emph{multus} plugin~\cite{multus}.
User equipment have \gls{IP} addresses in a dedicated \texttt{/16} network, split into two \texttt{/17} networks for the Paris and Lyon regions.
Calico only manages the \gls{IP} host routes for the Kubernetes pod network, so we manually add \gls{IP} routes between all user equipment and the data network to all cluster nodes.
Our setup enforces physical separation by mapping central and remote deployments to specific nodes with labels.
One node is a dedicated Kubernetes control plane node, one node maps to the central region, and six nodes each map to Paris and Lyon.
We provision the NetMob traffic data on all nodes, so user equipment and application server pods can mount it as volumes.
% In addition, user equipment and application server pods mount a volume to log the evaluation results.
We fix the scheduling of \gls{SSF} pods with node selectors, so that they persistently use the same volume across restarts.

\subsection{Workloads}
We use \emph{Helm}~\cite{helm} to manage the workloads deployed on the cluster.
We base our deployments on the excellent Helm chart collections of \emph{Gradiant}~\cite{gradiant} and \emph{Bitnami}~\cite{bitnami}.
We install cores based on \emph{Open5GS}~\cite{open5gs} with additional pods for our \gls{SSF} and \gls{CMF}, using custom Helm values for raft, active, and failover cores.
We install base stations and user equipment using \emph{ueransim}~\cite{ueransim} with a custom \texttt{ueransim-ues} Docker image that injects our evaluation code into all user equipment pods.
In addition, we develop custom Helm releases for our \gls{SSF} and \gls{CMF} network functions, application servers, and load balancers.

\paragraph{Core Network}
In the central region, we deploy six \emph{active} cores, five of which have Raft functionality (\emph{raft} cores).
Active cores operate all \gls{5G} network functions and use the \gls{SSF} to replicate state across the country.
Raft cores are the voting members of the Raft-based MongoDB replica set~\cite{mongodb_2025_replicaset}, while active and failover cores are non-voting members.
We deploy two \emph{failover} cores:
The one in Lyon connects to one of the central raft cores, and the one in Paris connects to a central (non-raft) active core.
The only network functions of failover cores operating non-stop are \gls{SSF} and \gls{CMF}.

\paragraph{Connectivity Monitoring and State Synchronization}
The \gls{CMF} of the core replicas can have one of two roles: \emph{active} or \emph{passive}.
Active \glspl{CMF} expose a \texttt{/health} endpoint to be probed by passive \glspl{CMF}.
They also expose a \texttt{/state} endpoint that regional application deployments can probe to monitor whether the region is in normal operation or island operation.
In normal operation, the central \gls{UPF} covers the traffic from all regions.
Once the \gls{CMF} detects that the configured active core is unavailable, it signals the transition to island operation via its \texttt{/state} endpoint.
Then, the failover core activates as an island core that operates all \gls{5G} network functions based on the most recent state synchronized through the \gls{SSF}.
Finally, the regional \glspl{UPF} take over managing the traffic from the regional \texttt{/17} user equipment networks.

\paragraph{Access Network}
We deploy 176 base stations in Paris and 134 base stations in Lyon based on \texttt{ueransim-gnb}.
Each base station is configured to connect to an \gls{AMF} via the N2 interface and to an \gls{UPF} via the N3 interface.
In normal operation, all base stations connect to the central active core, but upon transition into island operation, they reconnect to the regional \gls{AMF} and \gls{UPF}.
We map each NetMob tile to its closest base station and combine the tiles mapped to the same base station in one \texttt{ueransim-ues} pod, resulting in a mean number of \num{6.02} tiles representing the users of \SI{602}{\meter\squared} per user equipment pod.
The Paris region is denser than the Lyon region, so the cell size operated by one base station in Paris is smaller than in Lyon.
Each user equipment pod has a custom network interface that tunnels user traffic through the base station to the \gls{5G} core, which then forwards it to the data network via the N6 interface.

\begin{emphbox}
    \textbf{Large-Scale \gls{5G} Connectivity Simulation.}
    We simulate a core network with eight \gls{5G} cores, an access network with \numgnbs base stations and the users of a combined area of \SI{18.71}{\kilo\meter\squared} on a 14-node Kubernetes cluster.
    Our setup establishes \gls{IP} connectivity between users and application servers via the \gls{5G} user-plane protocol stack.
    State-of-the-art \gls{5G} cores become island-ready with our \gls{CMF} and \gls{SSF}.
\end{emphbox}

Our setup enables various \gls{5G} connectivity experiments, \eg, testing application-layer connections on top of \gls{IP} connectivity between users and the simulated Internet.

% ################################################################################
% ################################################################################
% ################################################################################
% ################################################################################
% ################################################################################

\section{Evaluation}
\label{sec: pub-d evaluation}

We run experiments on the Kubernetes cluster to test practical aspects of our \gls{SSF} and \gls{CMF} implementation and answer RQ2 regarding the added overhead.
Each experiment uses the traffic of one NetMob timeslot and simulates application traffic exchanged between users and application servers via the \gls{5G} user plane.
NetMob has a temporal resolution of 15-minute timeslots, collected over 77 consecutive days, resulting in \num{7392} total timeslots.
Since these timeslots contain an unbalanced mix of working and non-working days, and we abstract from day-night differences, we randomly select timeslots for the simulation and leverage the large number of data points to run our experiments multiple times for higher data quality.

\subsection{Implementation}

\begin{figure}
    \includegraphics[width=\linewidth]{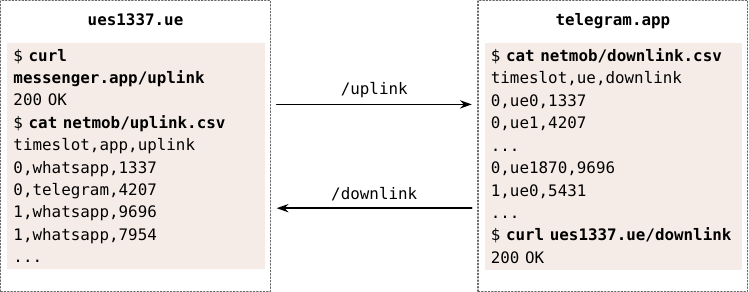}
    \caption[Evaluation API Endpoints]{%
        \captionheadline{Evaluation \gls{API} Endpoints}
        User equipment and application servers exchange NetMob uplink and downlink traffic through \Gls{5G}.
    }
    \label{fig: pub-d eval traffic}
\end{figure}

\Cref{fig: pub-d eval traffic} depicts the uplink and downlink traffic flow based on the NetMob dataset.
User equipment have a dedicated evaluation endpoint \texttt{/downlink}, on which they listen for traffic, and application servers have equivalent \texttt{/uplink} endpoints.
\emph{Prometheus}~\cite{prometheus} monitors the Kubernetes cluster, allowing an exhaustive view of the system’s resource consumption during simulation.
%  and \emph{Grafana}~\cite{grafana} visualizes the collected metrics.
In the data network, we deploy WhatsApp and Telegram application servers implemented in \emph{Go} that simulate WhatsApp with a central backend architecture (\ie, a single server in the central region) and Telegram with an island-ready backend architecture (\ie, servers in all regions).
We put a load balancer in front of each application backend to make the architecture transparent for the user equipment.

\subsection{Switching Latency}

\begin{figure}
    \includegraphics[width=\linewidth]{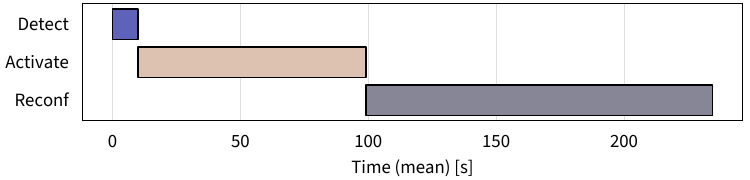}
    \caption[Switching Latency]{%
        \captionheadline{Switching Latency}
        Reconfiguring the access network accounts for more than half of the transition time.
    }
    \label{fig: pub-d switching latency}
\end{figure}

The \gls{CMF} waits for $T_F$ failed heartbeats before signaling the transition to island operation.
High switching thresholds lead to a high number of lost packets before connectivity is restored, while low thresholds risk false-positive transitions to island operation.
Finding a suitable threshold is therefore crucial for a smooth operation.
We simulate normal operation before isolating the Paris region from global connectivity at $t_0$.
It takes three steps to transition from global to island connectivity:
\emph{Detect} $T_F$ failed heartbeats and start the transition,
\emph{activate} the regional island core, and
\emph{reconfigure} the regional access network.
Choosing a small $T_F$ can lead to false switches, as we observe for $T_F\leq3$.
Next, we measure the transition time until island connectivity is established:
After the final reconfiguration step, island connectivity is established at $t_1$, and the total switching latency is $\lambda=t_1 - t_0$.
\Cref{fig: pub-d switching latency} depicts the mean duration per transition step in a broken bar chart.
The results indicate that the switching threshold $T_F$ has a limited impact on switching latency, as activating the island core and reconfiguring the access network account for most of the delay.
Our findings suggest a conservative $T_F$ to avoid false-positive transitions, and to assess acceleration options for the activation and reconfiguration steps.

\subsection{Synchronization Overhead}

\begin{figure}
    \includegraphics[width=\linewidth]{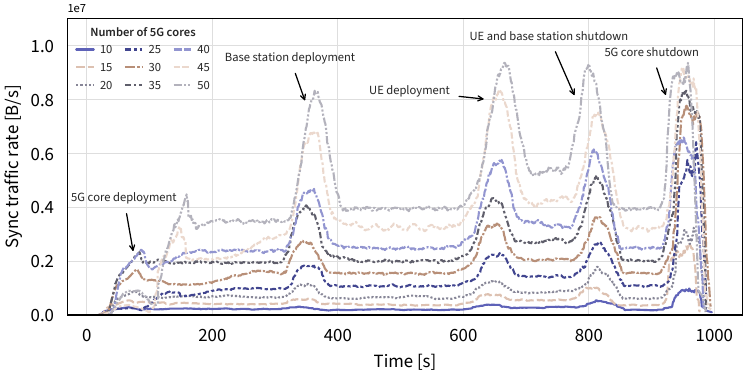}
    \caption[Synchronization Overhead]{%
    \captionheadline{Synchronization Overhead}
    The total synchronization traffic grows faster than linearly with the number of \gls{5G} cores.
    }
    \label{fig: pub-d sync overhead}
\end{figure}

In this experiment, we aim to study the message overhead introduced by the \gls{SSF} in the \gls{5G} core network.
We deploy a \emph{MongoDB exporter}~\cite{percona} sidecar container to each \gls{SSF} pod to record the overhead of the MongoDB replica set.
\Cref{fig: pub-d sync overhead} depicts the total synchronization traffic rate in the cluster over time.
In the experiment, we deploy the \gls{5G} cores at $t_0=\SI{0}{\second}$ and let them run idle until, at $t_1=\SI{300}{\second}$, we deploy the 310 base stations of Paris and Lyon.
This leads to the first traffic spike around $t=\SI{260}{\second}$, followed by another idle period until $t_2=\SI{600}{\second}$.
Then, we deploy the 310 user equipment, triggering subscriber state updates that must be synchronized cluster-wide, leading to the second peak at $t=\SI{640}{\second}$.
The third traffic peak stems from the shutdown of all user equipment and base stations, and the final peak from the shutdown of the \gls{5G} cores.

Our results indicate that the mean synchronization traffic increases from \SI{273}{\kilo\byte\per\second} to \SI{4.54}{\mega\byte\per\second} as the number of cores increases from 10 to 50 in a country-wide \gls{5G} deployment.
For the 50-core deployment, the peak synchronization rate is \SI{9.36}{\mega\byte\per\second} for \numtiles subscribers.
More generally, the measured synchronization overhead increases faster than linearly with the number of \gls{5G} cores, which is consistent with the additional replication and consensus traffic of the MongoDB replica set.

% ################################################################################
% ################################################################################
% ################################################################################
% ################################################################################
% ################################################################################

\section{Discussion}
\label{sec: pub-d discussion}

Our goal was to integrate state synchronization and connectivity monitoring functionality into state-of-the-art \gls{5G} cores to make them island-ready.
To this end, we studied two research questions concerned with the applicability of consensus and failure detection protocols to the \gls{SSF} and \gls{CMF} (RQ1) and the overhead added to the core network (RQ2).

\begin{emphbox}
    \textbf{Answers to RQ1 and RQ2.}
    Heartbeat-based failure detection and Raft-based consensus are ideal choices to implement connectivity monitoring in the \gls{CMF} and state synchronization in the \gls{SSF}.
    Our findings indicate that the total synchronization overhead to enable island readiness grows faster than linearly with the number of \gls{5G} cores.
\end{emphbox}

% FUTUREWORK ???
% Heartbeat-based failure detection is an ideal solution for our use case due to its lightweight implementation with minimal overhead.
% Still, future work might investigate the design and implementation of \gls{CMF} with more advanced failure detection protocols, enabling new use cases, such as the dynamic formation of islands beyond preconfigured hierarchies.
% Suitable starting points include \emph{Gossip}~\cite{renesse_2009_gossip} or \emph{memberlist}~\cite{hashicorp_2025_memberlist}, which have received much attention in related work.

\subsection{POBTOG Topologies}
We worked with a progressive \gls{5G} deployment regarding its island readiness with $N_0+N_1=30$ cores in normal operation and $N_2=500$ failover \gls{5G} cores for island operation.
In particular, $N_2$ determines the size of island-ready regions because the topology and users connected to the same failover core are part of the region, which the failover core can provide island connectivity to.
Still, expecting \glspl{MNO} to operate 500 failover cores per country is ambitious, and we encourage future work to study island readiness in settings with fewer larger island-ready regions. 
For instance, considering all infrastructure connected to the same active core would result in $N_1=25$ regions for France around the green squares in \Cref{fig: pub-d simulation france}.

\subsection{NetMob Traffic Data}

We integrated POBTOG topologies with the NetMob dataset, which was available through the NetMob challenge, for our evaluation.
Alternatively, \eg, without a suitable traffic dataset at hand, future work could configure POBTOG to generate an additional layer $L_4$ and choose a smaller $P$ suitable to let $N_4$ match the number of cellular users per tile.

Since NetMob reports normalized traffic and prohibits attempts to reverse the normalization, we used a scaling factor $S$ that multiplies the traffic values to determine the number of packets that users and application servers send through the network per 15-minute timeslot.
Our intention is that a suitable $S$ can align our simulation with the real world:
We assume the highest recorded traffic value to correspond to the capacity limit $C_\text{real}$ of the French cellular network, so NetMob data implicitly records the normalized real-world capacity limit $C_\text{norm}$.
Therefore, identifying an $S$ that satisfies $S\cdot C_\text{norm}\approx C_\text{sim}$ allows simulating realistic traffic workloads without compromising the privacy of the collected NetMob data.
To this end, we test increasing traffic alignment factors $S$ until our cluster reaches its capacity limit.
We found that the \gls{RRC} connection between the user equipment and the base station is the first point of failure.
This error occurs when a base station is overloaded by uplink requests from the user equipment and discards the user equipment's \gls{RRC} setup request, assuming that the user equipment context already exists.
As a result, user equipment cell selection fails, and the user equipment cannot connect to the data network.
% \Cref{fig: pub-d alignment} shows the average uplink and downlink packet delivery rate (PDR) and the number of RRC connection failures for increasing normalization factors $S$.
The results show that the base stations in our setup are overloaded for $S\geq 0.03125$, suggesting that $S=.001$ aligns our setup with the real world.
Further experiments are necessary to generalize this finding, which we encourage for future work.

% ################################################################################
% ################################################################################
% ################################################################################
% ################################################################################
% ################################################################################

\section{Conclusion}
The main contributions of this work are the introduction and evaluation of state synchronization and connectivity monitoring functionality for island-ready \gls{5G} cores.
To this end, we integrate a Raft-based \gls{SSF} and a heartbeat-based \gls{CMF} to the \gls{5G} control plane (RQ1), ensuring island readiness.
We generate a nationwide \gls{5G} network topology for France, on which we conduct a large-scale simulation study using a real-world cellular traffic dataset.
We orchestrate an island-ready \gls{5G} core network with eight cores and an access network of \numgnbs base stations on a 14-node Kubernetes cluster.
Our setup enables \gls{IP} connectivity between all \gls{5G} user equipment and the simulated data network.
We conduct experiments to study the transition latency of the \gls{CMF} and the synchronization overhead of the \gls{SSF} (RQ2), finding that most transition delay stems from the reconfiguration of the access network, and that synchronization overhead grows more than linearly with the number of \gls{5G} cores.
With our contributions, we hope to pave the way for the widespread realization of island readiness.

% ################################################################################
% ################################################################################
% ################################################################################
% ################################################################################
% ################################################################################

\section*{Acknowledgments}
This work was supported by the Federal Ministry of Research, Technology and Space of Germany in the project Open6GHub+ (grant number: 16KIS2407) and by the LOEWE initiative (Hesse, Germany) within the emergenCITY center [LOEWE/1/12/519/03/05.001(0016)/72]. Parts of our work used data made available by Orange within the Netmob 2023 Challenge~\cite{martinez_2023_netmob}.

\printbibliography

\end{document}